\documentclass[%
 aip,
 jap,%
 amsmath,amssymb,
reprint,%
]{revtex4-1}

\draft

\usepackage{graphicx}% Include figure files
\usepackage{sidecap}

\usepackage{dcolumn}% Align table columns on decimal point
\usepackage{bm}% bold math
\usepackage{chemfig} 
\usepackage{miller}

\usepackage{siunitx} % to get proper distances between num and unit 
\usepackage{xspace}
\usepackage{xargs} % Use more than one optional parameter in a new commands
\usepackage{multirow}
\usepackage{nicefrac}
\usepackage[colorinlistoftodos,prependcaption,textsize=tiny]{todonotes}
\newcommandx{\unsure}[2][1=]{\todo[linecolor=red,backgroundcolor=red!25,bordercolor=red,#1]{#2}}
\newcommandx{\change}[2][1=]{\todo[linecolor=blue,backgroundcolor=blue!25,bordercolor=blue,#1]{#2}}
\newcommandx{\info}[2][1=]{\todo[linecolor=gray,backgroundcolor=gray!25,bordercolor=gray,#1]{#2}}
\newcommandx{\improvement}[2][1=]{\todo[linecolor=Plum,backgroundcolor=Plum!25,bordercolor=Plum,#1]{#2}}
\newcommandx{\thiswillnotshow}[2][1=]{\todo[disable,#1]{#2}}

\usepackage{hyperref}% add hypertext capabilities

\newcommand{\ccn}{$\left[\text{C}_2\text{N}\right]_\text{As}$\xspace}
\newcommand{\cnn}{$\left[\text{CN}_2\right]_\text{As}$\xspace}
\newcommand{\cn}{$\left[\text{CN}\right]_\text{As}$\xspace}

\makeatletter
\newcommand\xleftrightarrow[2][]{%
  \ext@arrow 9999{\longleftrightarrowfill@}{#1}{#2}}
\newcommand\longleftrightarrowfill@{%
  \arrowfill@\leftarrow\relbar\rightarrow}
\makeatother
\begin{document}

\preprint{Special Topic on Defects in Semiconductors}

\title{Symmetry and structure of carbon-nitrogen complexes in gallium arsenide from infrared spectroscopy and first-principles calculations}% 

\author{Christopher K\"unneth}
\email{kuenneth@hm.edu}
%\altaffiliation[Also at ]{Physics Department, XYZ University.}%Lines break automatically or can be forced with \\
\author{Simon K\"olbl}%
%\email{rmaterli@hm.edu}
\author{Hans Edwin Wagner}
\affiliation{%
    Munich University of Applied Sciences, Department of Applied Sciences and Mechatronics, Lothstr. 34, D-80335 Munich, Germany
}%
\author{Volker H\"aublein}
\affiliation{%
  Fraunhofer Institute for Integrated Systems and Device Technology IISB, Schottkystrasse 10, 91058 Erlangen, Germany
}%

\author{Alfred Kersch}%
\email{akersch@hm.edu}
\author{Hans Christian Alt}%
\email{hchalt@hm.edu}
\affiliation{%
    Munich University of Applied Sciences, Department of Applied Sciences and Mechatronics, Lothstr. 34, D-80335 Munich, Germany
}%

\date{\today}% It is always \today, today,
             %  but any date may be explicitly specified

\begin{abstract}
Molecular-like carbon-nitrogen complexes in GaAs are investigated both experimentally and theoretically. Two characteristic high-frequency stretching modes at \num{1973} and \SI{2060}{cm^{-1}}, detected by Fourier transform infrared absorption (FTIR) spectroscopy, appear in carbon- and nitrogen-implanted and annealed layers. From isotopic substitution it is deduced that the chemical composition of the underlying complexes is CN$_2$ and C$_2$N, respectively. Piezospectroscopic FTIR measurements reveal that both centers have tetragonal symmetry. For density functional theory (DFT) calculations linear entities are substituted for the As anion, with the axis oriented along the \hkl<100> direction, in accordance with the experimentally ascertained symmetry. The DFT calculations support the stability of linear N-C-N and C-C-N complexes in the GaAs host crystal in the charge states ranging from $+3$ to $-3$. The valence bonds of the complexes are analyzed using molecular-like orbitals from DFT. It turns out that internal bonds and bonds to the lattice are essentially independent of the charge state. The calculated vibrational mode frequencies are close to the experimental values and reproduce precisely the isotopic mass splitting from FTIR experiments. Finally, the formation energies show that under thermodynamic equilibrium CN$_2$ is more stable than C$_2$N.

\end{abstract}

%\pacs{Valid PACS appear here}% PACS, the Physics and Astronomy
                             % Classification Scheme.
 %Use showkeys class option if keyword
                              %display desired
\maketitle

\section{Introduction}

Carbon and nitrogen are important light-element impurities in GaAs bulk crystals as well as in GaAs-based epitaxial layers. Carbon as a common substitutional acceptor (C$_\text{As}$) is used in semi-insulating GaAs crystals to stabilize the Fermi level near the mid gap position \cite{Jurisch2005}. Carbon is also widely employed to grow highly p-type doped epitaxial layers \cite{Saito1988,Abernathy1989}. In the last couple of decades, the behavior of nitrogen in GaAs has attracted much interest in the context of dilute nitrides. The ternary system GaAs$_{1-x}$N$_x$, where a small amount $x$ of the anion sites is replaced by the isovalent N atom, is in the focus of intense research. Addition of nitrogen to GaAs during epitaxial growth leads to a remarkable shrinkage of the band gap with an initial slope of about \SI{180}{\milli\electronvolt} for \SI{1}{\percent} ($x = 0.01$) of nitrogen \cite{Weyers1992}. This opens up the possibility of band gap engineering for example for highly efficient multi-junction solar cells \cite{Jackrel2007}. Furthermore, the reversal of the band gap shrinkage by hydrogen passivation has been observed and studied theoretically \cite{Polimeni2002}.

Complexes of carbon and nitrogen have an influence on these doping-related material properties. The first report of such a complex was given by Ulrici and Clerjaud \cite{Ulrici2005a} in 2005. They observed a sharp local vibrational mode (LVM) at \SI{2087.1}{\centi\metre^{-1}} at \SI{7}{\kelvin} in the related semiconductor gallium phosphide. From the detection of small satellite bands caused by the natural isotopes $^{13}$C and $^{15}$N, the assignment to a CN complex was straightforward. Furthermore, uniaxial stress measurements showed that the defect has tetragonal symmetry with the carbon-nitrogen bond aligned along the \hkl<100> axis of the crystal. They give also a short note on a band at \SI{2088.5}{cm^{-1}} in one particular GaAs crystal doped with both carbon and nitrogen. From the nearly identical vibrational frequency and similar temperature dependence as in GaP case they tentatively suggest an analogous defect in GaAs. However, in the course of this and the extensive previous study on bulk GaAs crystals this band was not observed \cite{Alt2013,Alt2015}. 

Another carbon-related band in GaAs appears at \SI{2059.6}{cm^{-1}} at \SI{7}{\kelvin} \cite{Ulrici2005}. It was first tentatively attributed to a center involving carbon and oxygen, but its origin is a carbon-nitrogen complex. The band can be observed in carbon-rich GaAs crystals after long-term annealing at around \SI{700}{\degree C}. The intensity of the band increases with the carbon concentration as determined by the C$_\text{As}$ LVM at \SI{582}{cm^{-1}}. The participation of carbon in the complex is ascertained by the existence of a small $^{13}$C satellite at \SI{2003.8}{cm^{-1}}. From its relative intensity, the incorporation of precisely one C atom can be inferred. Furthermore, from the strength of the main band and the decrease of the substitutional carbon concentration, it can be concluded that a considerable fraction of the carbon impurities is transformed to this complex during annealing \cite{Ulrici2005}. 

Contrary to carbon, the role of nitrogen in this complex was more difficult to assess. Nitrogen as a substitutional isoelectronic impurity (N$_\text{As}$) in bulk GaAs can only be detected by mass spectrometry or LVM absorption for concentrations above \SI{e15}{cm^{-3}} not far from the maximum doping level of some \SI{e16}{cm^{-3}} \cite{Alt1997}. Usually, the nitrogen contamination coming from the pyrolytic boron nitride crucible and the nitrogen atmosphere during crystal growth is much lower. The verification of nitrogen incorporation was carried out by some of the present authors using GaAs samples implanted with $^{12}$C and the nitrogen isotopes $^{14}$N and $^{15}$N \cite{Alt2013}. In this case, high concentrations of carbon-nitrogen complexes responsible for the \SI{2060}{cm^{-1}} absorption band are generated in a near-surface layer. The isotope splitting in samples implanted with both $^{14}$N and $^{15}$N allows the identification of the chemical composition of the complex as \cnn. Furthermore, in the same layer structures a second LVM band at \SI{1973}{cm^{-1}} was found originating from another defect related to carbon and nitrogen. In contrast to \cnn, this second center contains only one N atom. So far, no further information was available.

\begin{SCfigure*}
\includegraphics[width=.62\paperwidth]{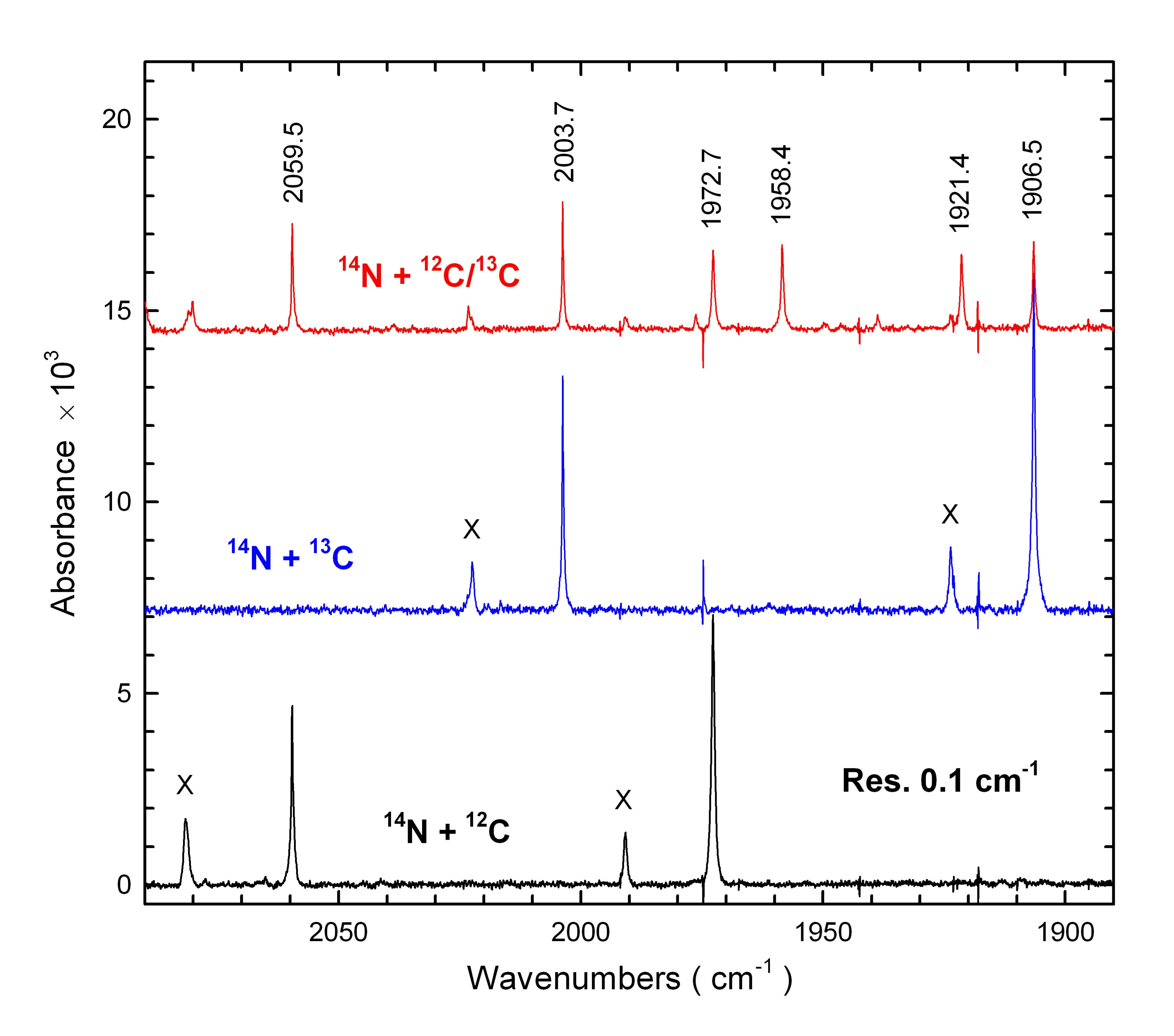}
\caption{\label{fig:isotope_shift} Infrared absorption spectra of carbon-nitrogen related vibrational modes in GaAs at \SI{9}{\kelvin} after implantation of $^{14}$N and $^{12}$C and/or $^{13}$C and annealing at \SI{700}{\degree C}. Spectra are baseline-corrected and shifted vertically for clarity. Bands marked by 'X' appear only in implanted samples and are most probably due to defects involving residual irradiation damage.}
\end{SCfigure*}
Carbon-nitrogen complexes in GaAs and GaP were studied theoretically by Limpijumnong et al.\cite{Limpijumnong2008} based on density functional theory (DFT) calculations. Considering the carbon-nitrogen molecule in different charge states they found that in general the substitutional configuration on the anion site $\left[\text{CN}\right]_\text{As}$ is energetically more favorable than the interstitial one $\left[\text{CN}\right]_\text{i}$. The stretching frequency has a strong dependence on the charge state of the molecule. Only the $2+$ charge state has a frequency high enough to be compatible with the experimental value found in GaP\cite{Ulrici2005a}. However, this charge state relaxes from the substitutional site with tetragonal symmetry to a low-symmetry off-center position contradicting the experimental result.

The present work completes previous FTIR studies by piezospectroscopic experiments for both centers in samples with surface-implanted layers (Subsections  \ref{subsec:exp_isotopes} and \ref{subsec:exp_stress}). In Subsection \ref{subsec:theo_vibrational} experimental results are used as starting configurations for extensive DFT calculations and analyses of the LVM frequencies. In Subsection \ref{subsec:theo_molecular} a Kohn-Sham (KS) orbital interpretation of the related eigenstates is presented and in Subsection \ref{subsec:theo_formation} the formation energies of the \cnn and \ccn defect.

\section{Experimental and computational methods}\label{sec:experimental}
Single-crystalline GaAs containing carbon-nitrogen complexes in high concentration in a near-surface layer was produced by implantation of carbon and nitrogen into \SI{5}{mm} thick semi-insulating wafers with \hkl{100} or \hkl{110} surface orientation. The total implantation dose was \SI{5e15}{cm^{-2}} for both nitrogen and carbon. Different ion energies up to \SI{200}{keV} were used to obtain a roughly uniform concentration depth profile. FTIR samples were cut from this material and treated by rapid thermal annealing (RTA) at \SI{700}{\degree C} for \SI{60}{s} under nitrogen atmosphere. Infrared (IR) absorption measurements were carried out with a Bruker Vertex 80v FTIR vacuum spectrometer equipped with a global source, a KBr beam splitter and a MCT detector. The samples were cooled to \SI{77}{\kelvin} or \SI{9}{\kelvin} in an optical cryostat. The implanted layer was aligned perpendicular to the IR light beam in order to get a maximum signal from the carbon-nitrogen centers. For piezospectroscopic investigations, samples with approximate dimensions of $20 \times 5 \times 5$ \SI{}{mm^{3}} were prepared with the long axis oriented along the main crystallographic directions \hkl<100>, \hkl<110>, or \hkl<111>. Uniaxial stress was applied to the sample along the long axis by a push rod coupled to a pneumatic cylinder using a home-made apparatus. The force on the sample was calculated from the gas pressure in the cylinder that was monitored on a precision gauge.

Numerical results were obtained with the all-electron DFT code FHI-Aims \cite{Blum2009,Knuth2015,Marek2014,Auckenthaler2011,Havu2009} which utilizes numerical atom-centered basis function. The LDA (PW\cite{Perdew1992} parameterization) or HSE06\cite{Ren2012a} approximation with the mixing parameter $\alpha = 0.25$ and $\omega = 0.11 a_0^{-1}$ was chosen as the exchange-correlation functional. All calculations are based on a 64 atoms supercell of a GaAs crystal with the cubic F$\overline{4}3$m (No. 216) symmetry. Since FHI-Aims does not include symmetry considerations, all convergences were archived without symmetry constraints. The supercell of the defect structures \cnn ($\text{Ga}_{32} \text{As}_{31} \text{C}_1 \text{N}_2$) and \ccn ($\text{Ga}_{32} \text{As}_{31} \text{C}_2 \text{N}_1$) contains 66 atoms. A convergence study reveals that a k-point grid of $3 \times 3 \times 3$ with an electronic force convergence of \SI{5e-5}{eV\per\angstrom} with the tight basis set in the second tier is sufficient. The unit cell and ion positions were converged for all supercells up to \SI{5e-4}{eV\per\angstrom} except for charged cells where only ions were allowed to move keeping the unit cell as the neutral ones. Vibrational frequencies were carried out with the utility Phonopy \cite{Togo2015} using finite displacements. The structural relaxation and the vibrational frequencies in this work were calculated using LDA.  

Formation energies were calculated according to \cite{Freysoldt2014}
\begin{eqnarray}
E^\text{f}\left[X^q\right] &&= E_\text{tot}\left[X^q\right] - E_\text{tot}\left[\text{GaAs}\right] -\sum_i n_i \mu_i \nonumber\\ 
&& + q\left( E_\text{F} + E_\text{VBM}\left[\text{GaAs}\right] + \Delta V\left[X^0\right] \right) \nonumber\\
&&+ E_\text{corr}[X^q]
\label{eq:formation_energy}
\end{eqnarray}
with $E^\text{f}$ the formation energy, $E_\text{tot}$ the total energy, $n_i$ the numbers of impurities, $\mu_i$ the chemical potential of the impurity $i$, $E_\text{F}$ the Fermi level referenced to the energy of the valence band maximum $E_\text{VBM}$, $\Delta V$ the potential alignment, $E_\text{corr}$ the charge correction due to finite size of the unit cell and $ X \in \left \{ \left[\text{C}_2\text{N}\right]_\text{As}, \left[\text{CN}_2\right]_\text{As} \right\}$ the defect. Calculations for charged structures were carried out for the charges $q=-3,\ldots,+3$ for both defects with the cell fixed to the uncharged structure. For \cnn and \ccn the chemical potential were $\sum_i n_i \mu_i = -\mu_\text{As} + \mu_\text{C} + 2\mu_\text{N}$ and $\sum_i n_i \mu_i = -\mu_\text{As} + 2\mu_\text{C} + \mu_\text{N}$, respectively. $\mu_\text{As}$ was calculated from trigonal arsenic, $\mu_\text{C}$ from diamond and $\mu_\text{N}$ from a N$_2$ molecule. Figures of atomic structures and densities used in this publication were produced with VESTA \cite{Momma2011}.

%Analysis of defect-specific molecular valence orbitals and the bonding configuration in the host lattice was carried out exemplarily for $\text{CN}_2$ with a cluster calculation, using the Gaussian code \cite{Frisch2013} with the B3LYP functional and the 6-311 basis set. Relevant information is easily accessible with the powerful Gabedit \cite{Allouche2010} graphical interface tool. A spherical cluster $\text{(C}\text{N}_{2}\text{)Ga}_{16}\text{As}_{18}\text{H}_{36}$, covering the central $\text{CN}_2$ complex and Ga and As neighbors up to the fourth shell, was used. Dangling bonds were saturated by H atoms, taking into account different numbers of Ga and As atoms \cite{Korambath2000}. The resulting local $\text{CN}_2$ structures are practically identical to the more extensive FHI-Aims supercell calculations.  
 
\section{Results and Discussion}
\subsection{Isotope shifted satellite bands}
\label{subsec:exp_isotopes}

\begin{figure}
\includegraphics[width=0.36\paperwidth]{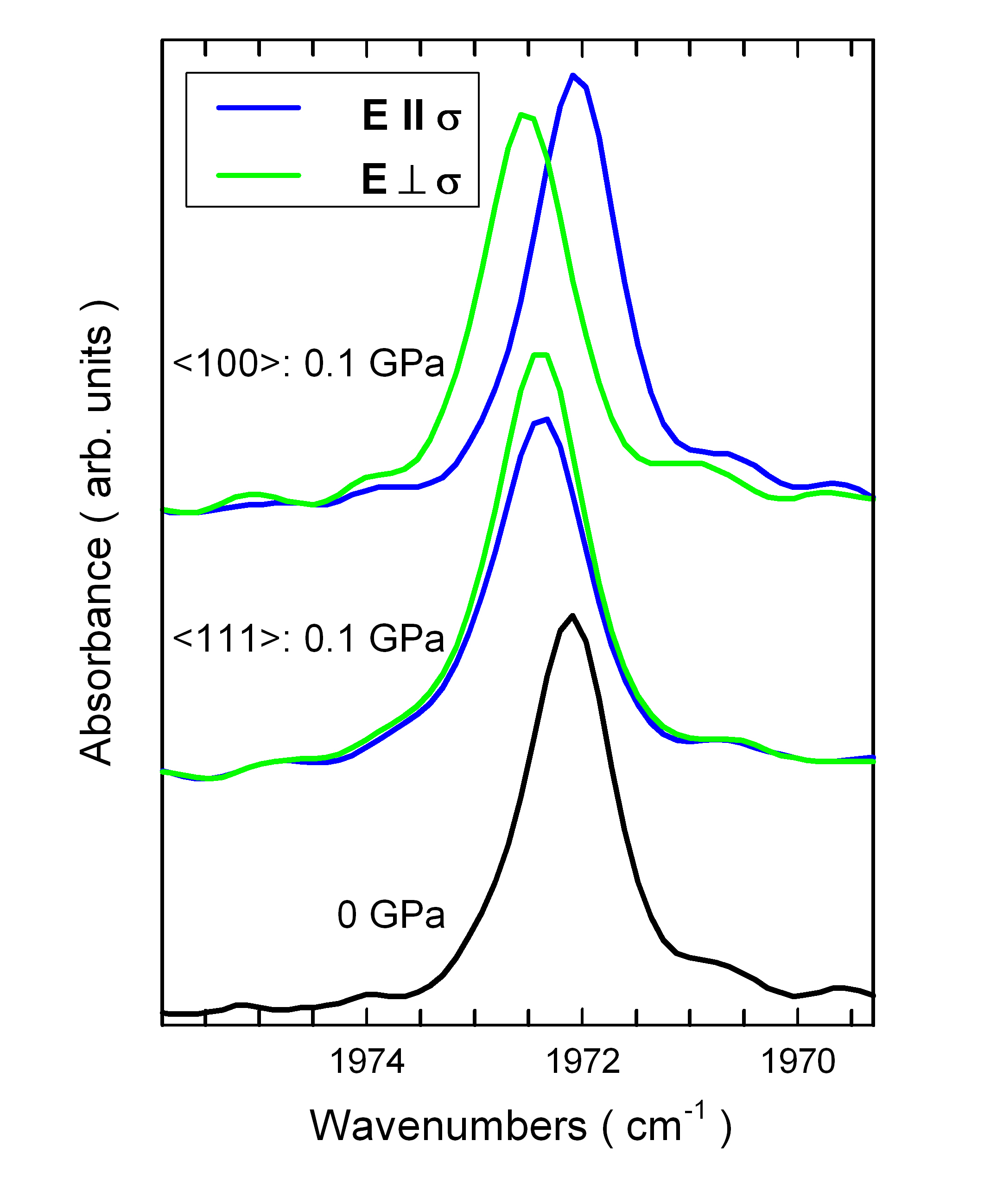}
\caption{\label{fig:absorption_splitting} Shift and splitting of the \ccn absorption band at \SI{1973}{cm^{-1}} band under stress in \hkl<100> and \hkl<111> direction at a sample temperature of 77 K.}
\end{figure}

The chemical composition of the carbon-nitrogen complexes giving rise to the absorption bands at \num{1973} and \SI{2060}{cm^{-1}} can be deduced from the characteristic mass-induced shift after isotopic substitution. As has been pointed out in Ref. 9, the replacement of $^{14}$N by $^{15}$N unambiguously leads to the conclusion that the former center contains one nitrogen atom, whereas the latter two in equivalent positions (configuration N-C-N). Here, we show the analogue experiment with the isotope pair $^{12}$C and $^{13}$C. Sample spectra at \SI{9}{\kelvin} after RTA at \SI{700}{\degree C} are shown in FIG. \ref{fig:isotope_shift}. 

Replacement of $^{12}$C by $^{13}$C shifts the \num{2060} band to \SI{2004}{cm^{-1}} and the \num{1973} band to \SI{1907}{cm^{-1}}\footnote{For simplicity we keep the terminology \num{1973} and \SI{2060}{cm^{-1}} band throughout the paper, although at \SI{77}{K} and in some implanted layers the peak position is closer to \num{1972} and \SI{2059}{cm^{-1}}, respectively}. The band at \SI{2004}{cm^{-1}} was first detected by Ulrici and Jurisch\cite{Ulrici2005} as a small satellite in carbon-doped bulk crystals caused by the naturally occurring $^{13}$C isotope (natural abundance \SI{1.1}{\percent}). It should be emphasized that for both bands the shift is larger than expected from a simple bi-atomic molecule. For the mixed-isotope implantation with $^{12}$C and $^{13}$C in the ratio 1:1, additional lines at \num{1921} and \SI{1958}{cm^{-1}} appear, see upper spectrum in FIG. \ref{fig:isotope_shift}. These lines belong to the \SI{1973}{cm^{-1}} defect and require that more than one C atom must be involved. The existence of four lines of equal intensity suggests that two C atoms on in-equivalent sites are incorporated called A and B. The simple statistical consideration that A and B are occupied with a probability of \SI{50}{\percent} by $^{12}$C or $^{13}$C brings about this result. Therefore, the chemical composition of the defect responsible for the \SI{1973}{cm^{-1}} band is \ccn (configuration C-C-N).
 
\begin{figure}
\includegraphics[width=0.45\textwidth]{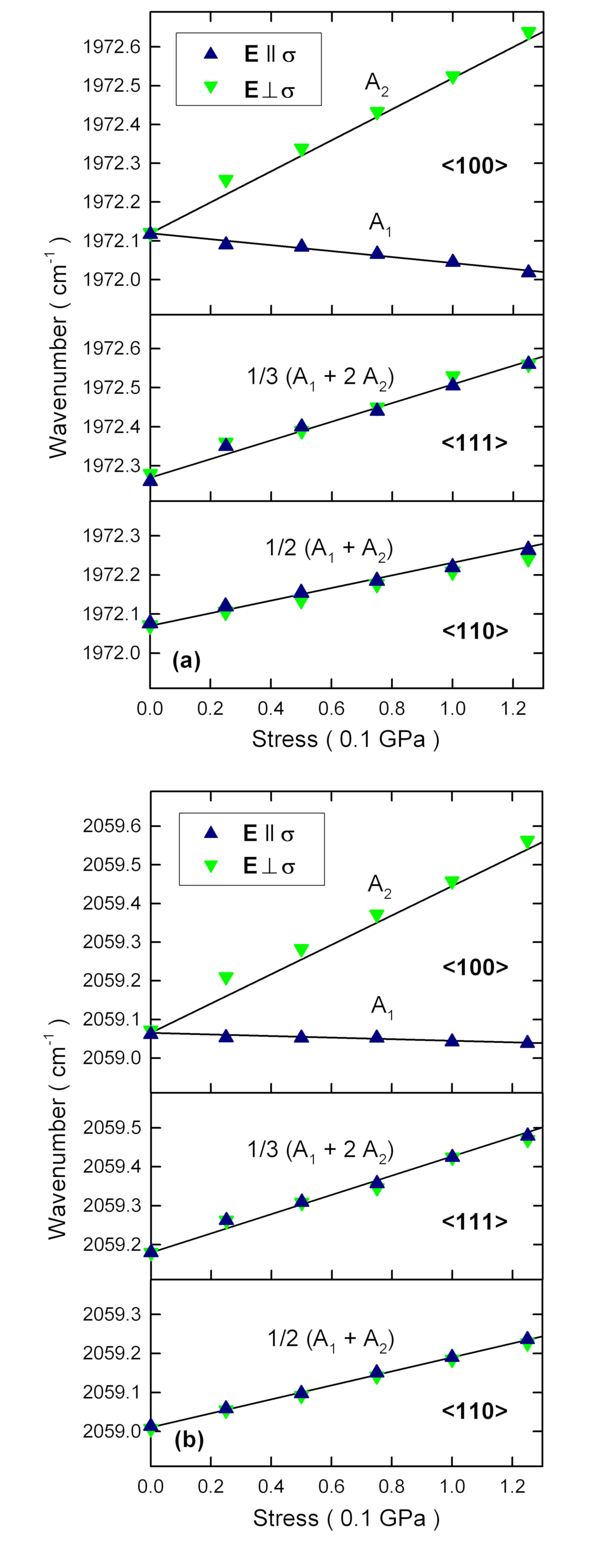}
\caption{\label{fig:different_pressures} Effect of uniaxial stress in \hkl<100>, \hkl<111>, and \hkl<110> directions on the frequency of the \ccn (a) and the \cnn band (b). Solid lines are fitted with piezospectroscopic parameters A$_1$ and A$_2$. Sample temperature is 77 K.}
\end{figure}

\subsection{Uniaxial stress experiments}
\label{subsec:exp_stress}

FIG. \ref{fig:absorption_splitting} illustrates the characteristics of the \SI{1973}{cm^{-1}} band under uniaxial stress. For stress along a \hkl<111> direction the band shifts to higher energy but does not split. In contrast, stress along the \hkl<100> direction splits the band into two components associated with a characteristic polarization behavior. One component (shifting to lower energy) is observed for polarization of the incident light parallel to the \hkl<100> stress direction whereas the other one (shifting to higher energy) appears for perpendicular polarization. This splitting and polarization behavior is typical for a tetragonal center in a cubic crystal and in full accordance with the systematic theoretical investigation of Kaplyanskii\cite{Kaplyanskii1964}. The splitting is caused by lifting off the orientational degeneracy. It simply reflects the fact that a tetragonal center with its primary alignment along one of the three equivalent \hkl<100> directions has two possible orientations relative to the stress direction in the \hkl<100> case, however only one for the \hkl<111> stress case.

\begin{table}
\caption{\label{tab:uniaxial_stress} Uniaxial stress characteristics of the \num{1973} and \SI{2060}{cm^{-1}} bands in GaAs.}
\begin{ruledtabular}
\begin{tabular}{ccccc}
Direction  & Piezospectr. & Intensity & \multicolumn{2}{c}{Slope (\si{cm^{-1}/GPa})} \\
of stress & parameter &  I$_\parallel : \text{I}_\perp$  & \SI{1973}{cm^{-1}} & \SI{2060}{cm^{-1}}\\
\hline
\multirow{2}{*}{\hkl<100>} & A$_1$ & $1:0$ & $-0.8$ & $-0.2$\\
                         & A$_2$ & $0:1$ & $4$ & $3.8$\\
\hkl<111> & $\nicefrac{1}{3}$(A$_1$ + $2\text{A}_2$) & $1:1$ & $2.4$ & $2.5$\\
\hkl<110> & $\nicefrac{1}{2}$(A$_1$ + $\text{A}_2$) & $1:1$ & $1.6$ & $1.8$\\
\end{tabular}
\end{ruledtabular}
\end{table}

A summary of the results of the uniaxial stress experiments are givin in FIG. \ref{fig:different_pressures} for \ccn in (a) and \cnn in (b). Measurements were carried out in the three crystallographic directions \hkl<100>, \hkl<111>, and \hkl<110> applying stress up to \SI{0.125}{GPa}. Polarized measuring light with the electric field vector $\vec{E}$ aligned parallel or perpendicular to the stress direction was used. For the \hkl<110> case, only one of the two branches was accessible due to the necessity of orienting the implanted layer perpendicular to the incident light (see section II). The piezospectroscopic parameters A$_1$ and A$_2$ according to Kaplyanskii's terminology \cite{Kaplyanskii1964} were derived from the fully polarized branches for \hkl<100> stress. In the case of the \SI{1973}{cm^{-1}} band, a complete fit of the experimental data is possible with A$_1 = $\SI{-0.8}{cm^{-1}\per GPa} and A$_2 =$\SI{4.0}{cm^{-1}\per GPa}. It should be mentioned that the peak position at zero stress shows some scattering between different samples, presumably due to residual strain in the implanted layer after annealing \cite{Alt2015}. The results for the \SI{2060}{cm^{-1}} band in (b) are shown for completeness. The piezospectroscopic parameters are within the experimental accuracy of about \SI{+-10}{\%} identical to the values obtained previously from bulk crystals \cite{Alt2013}. This emphasizes the fact that the \cnn center generated by implantation and RTA at \SI{700}{\degree C} is identical to the center observed in bulk crystals after long-term annealing at \SI{700}{\degree C} \cite{Ulrici2005}. Furthermore, the structural damage in the implanted layer and the high defect concentration obviously do not influence significantly the piezospectroscopic results. A compilation of all piezospectroscopic results is given in Table \ref{tab:uniaxial_stress}.

\begin{figure}
\includegraphics{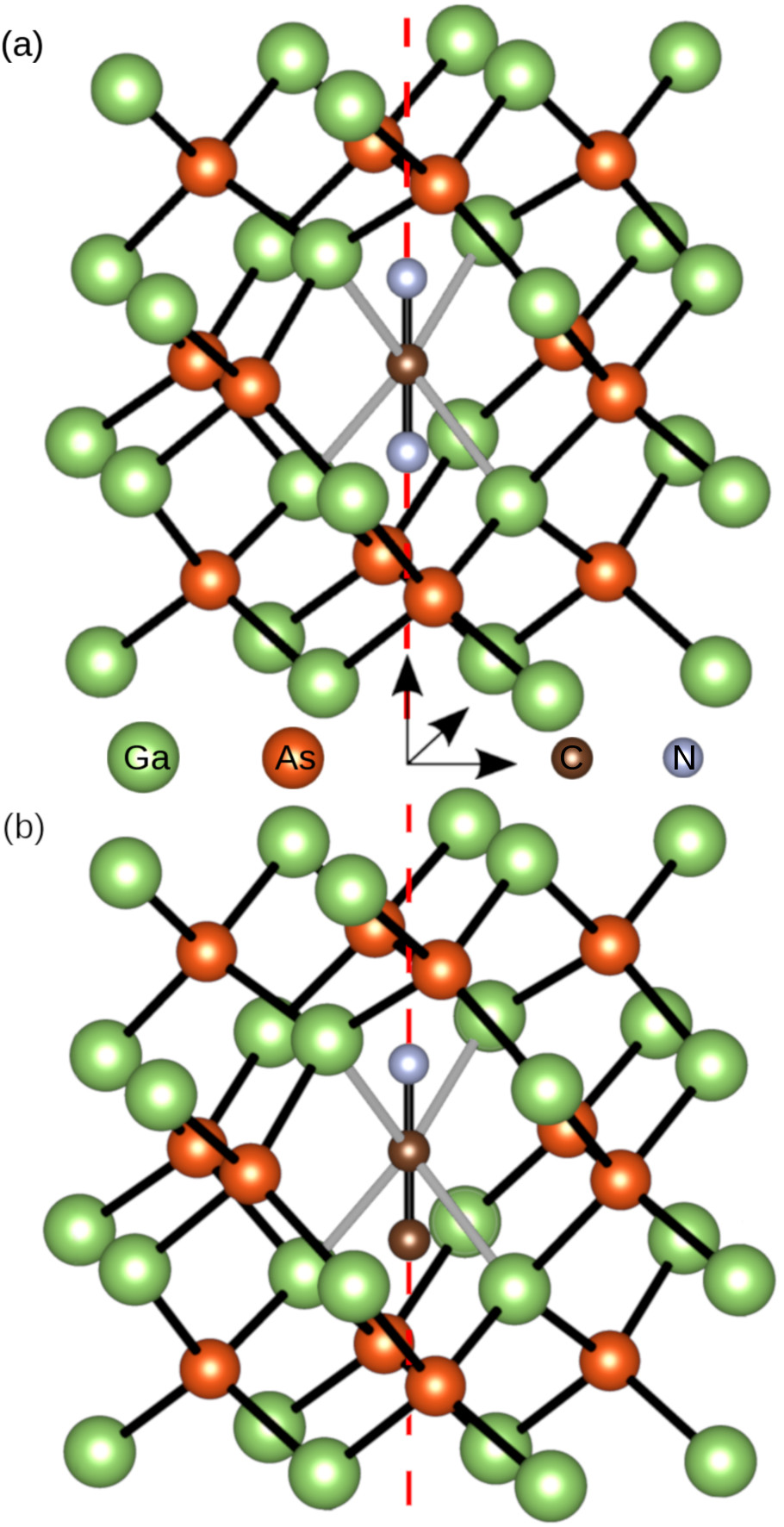}
\caption{\label{fig:structures} Incorporation of the \cnn (a) and the \ccn (b) complex in GaAs. Both complexes are aligned along the \hkl<100> axes of the GaAs host crystal. The gray bonds indicate the actual bonds for a pure GaAs crystal to the removed As. Some surrounding atoms are removed for better representation breaking the periodic supercell used in the calculations.}
\end{figure}

The behavior under stress is similar for both carbon-nitrogen complexes in GaAs not only qualitatively but also quantitatively. It is also worth mentioning that the size of the parameters A$_1$ and A$_2$ is near to the values obtained by Ulrici and Clerjaud \cite{Ulrici2005a} for the tetragonal carbon-nitrogen complex in GaP (\SI{2087}{cm^{-1}} band) indicating some similarities in bonding configuration. The negative sign of the parameter A$_1$ needs to be commented on. For \hkl<100> stress, this parameter reflects the response of the complex aligned with its axis parallel to the stress direction. Clearly, if the carbon-nitrogen bonds are compressed in this situation, the vibrational frequency should increase and A$_1$ should be positive. We will propose a model for this striking behavior in subsection \ref{subsec:theo_molecular}.

\subsection{Structural investigation and vibrational modes from DFT}
\label{subsec:theo_vibrational}

\begin{table}
\caption{\label{tab:phonon_both} Asymmetric stretching mode frequencies of the charged structures for $^{12}$C and $^{14}$N. $d_\text{CC}$ is the bond length between two carbon atoms and $d_\text{CN}$ between a carbon and a nitrogen atom.}
\begin{ruledtabular}
\begin{tabular}{c|cc|ccc}
& \multicolumn{2}{c|}{\cnn} & \multicolumn{3}{c}{\ccn} \\
Charge & $\omega$ & $d_\text{CN}$ & $\omega$ & $d_\text{CN}$ & $d_\text{CC}$\\
  (\si{\elementarycharge}) & (\si{cm^{-1}}) & (\si{\AA}) & (\si{cm^{-1}}) &  (\si{\AA}) &  (\si{\AA})\\
\hline
+2 & 2139.8 & 1.2204 & 2036.2 & 1.2403 & 1.2745 \\
+1 & 2139.3 & 1.2201 &    -   & 1.2418 & 1.2718 \\
 0 & 2128.3 & 1.2206 & 2049.2 & 1.2414 & 1.2705\\
-1 & 2127.4 & 1.2210 &    -   & 1.2418 & 1.2713 \\
-2 & 2122.3 & 1.2214 & 2042.0 & 1.2413 & 1.2718 \\
\end{tabular}
\end{ruledtabular}
\end{table}

Piezospectroscopy and its analysis strongly indicates that both complexes are aligned along the \hkl<100> directions of the GaAs crystal. The atomic configuration of the \cnn and the \ccn was chosen as N-C-N and C-C-N in a linear arrangement according to experimental results of subsections \ref{subsec:exp_isotopes} and \ref{subsec:exp_stress}. The structures after relaxation are depicted in FIG. \ref{fig:structures} for \ccn in (a) and \cnn in (b). The lattice constant amounts to \SI{5.605}{\AA} for the defect-free GaAs crystal in comparison to the experimental \SI{5.655}{\AA}. Although no symmetry constraints are applied to the DFT calculations, both defects are almost perfectly aligned along the \hkl<100> direction after the convergence is reached. Bond lengths of the complexes are documented in TAB. \ref{tab:phonon_both}.

\begin{figure}
\includegraphics[width=0.45\textwidth]{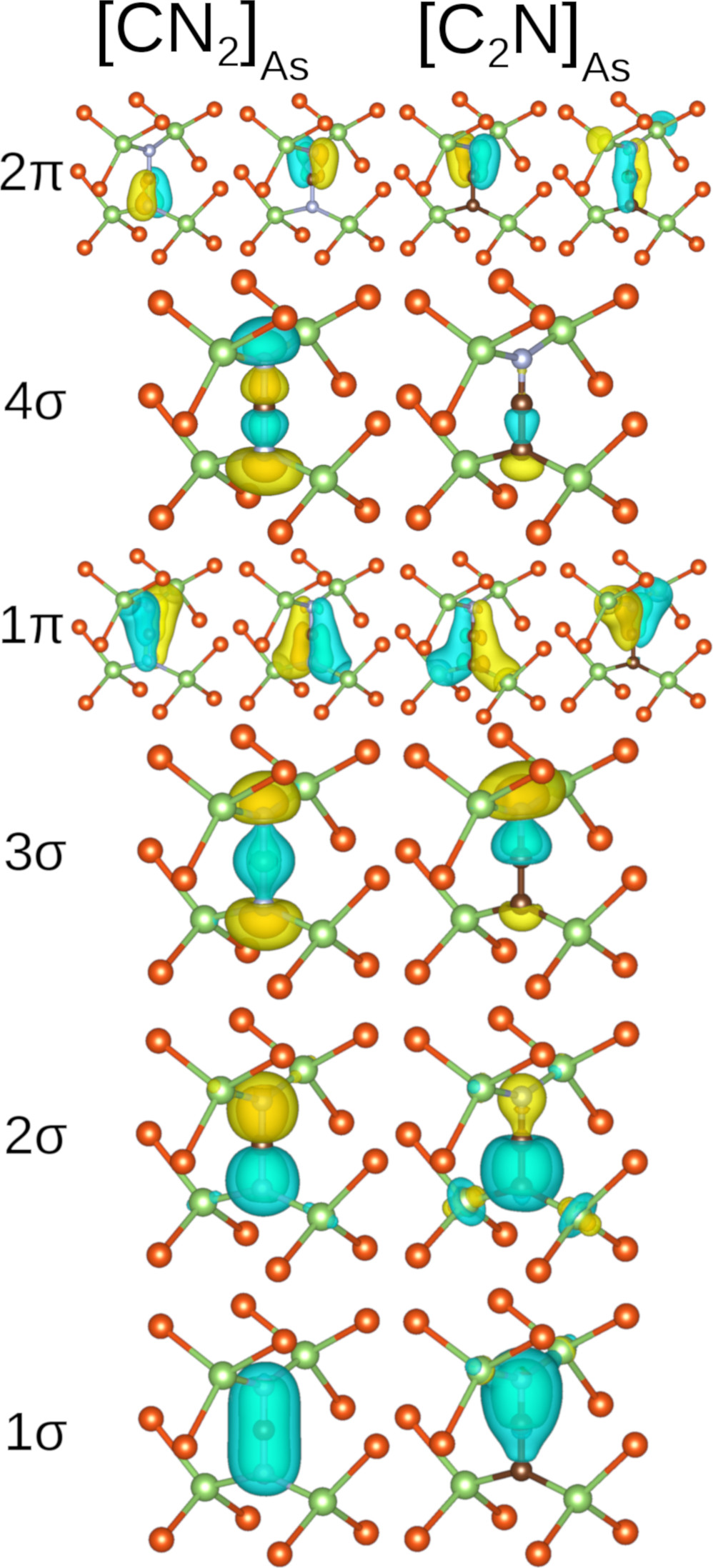}
\caption{\label{fig:molecularorbitals} Kohn-Sham molecular-like orbitals for \cnn and \ccn from HSE06 DFT calculations for different eigenvalues. The isosurfaces have the values \num{+-0.1} and \SI{+-0.05}{/a_0^{-3/2}} starting at the inner isosurface. Only strongly localized orbitals which are similar to the free molecule orbitals are shown.}
\end{figure}

Structures with a charge of $q=-3,\ldots,+3$ were calculated by adding and removing electrons from the supercells. In this way the effect of a modified Fermi level on the occupation of states and resulting change of bonding forces was calculated. Surprisingly, the charged \cnn and \ccn structures remained in a symmetric configuration with almost unchanged bond length, contrary to the case of the \cn complex as discussed by Limpijumnong et al.\cite{Limpijumnong2008}.  The calculation of the vibrational modes and their frequencies was carried out using the finite displacement\cite{Togo2015} method with a displacement of $\pm 0.001 \AA$ out of the equilibrium position for all atoms in the unit cell. To account for the effect of the lattice coupling on the LVM modes, we calculated the frequencies from the full dynamical matrix using the LDA approximation.

Inspection of the modes reveals that the highest frequency corresponds to the asymmetric and the second highest to the symmetric bond stretching mode. The third and fourth highest frequency are the two symmetry equivalent bending modes of the incorporated complex.\footnote{The third and fourth highest vibrational mode (bending) could not measured by FTIR experiments. Although the vibrational modes should be IR-active from symmetry, we assume that the oscillator strength is not high enough for detection.}

TAB. \ref{tab:phonon_both} shows the vibrational frequencies of the infrared active and asymmetric stretching modes for \cnn and \ccn in different charge states and the associated distance of the bond lengths. The change of the bond lengths is in the range of \SIrange{0.1}{0.01}{\%} accompanied by a similar small change of the frequency. An explanation for the small charge sensitivity of the vibrational modes is that electrons added or removed can not localize close to the carbon-nitrogen complex since no defect states are available in the band gap, see \ref{subsec:theo_formation} for further discussion. It should be noted, that the vibrational frequencies are calculated within the LDA approximation which is known to overestimate the delocalization of the electrons due to a self-interaction error. This error may decrease the variation of the bond length for different charges unphysically.

The vibrational frequencies for the carbon-nitrogen complexes in an isotopic configuration were calculated by substituting the masses in the dynamical matrix. Subsequently, the vibrational frequencies were obtained by diagonalization of the modified eigenvalue equation. TAB. \ref{tab:phonon_cn2} and \ref{tab:phonon_c2n} show the frequencies of the asymmetric stretching modes (highest frequency) for the \cnn and \ccn complexes in comparison to the assigned frequencies from the infrared measurements. To eliminate a systematic error of DFT and the neglect of anharmonic effects, a scaling factor $\sum_i \omega_{\text{calc}}\omega_{\text{exp}}/\sum_i \omega^2_{\text{calc}}$ for the frequency is introduced according to Ref. \onlinecite{Irikura2005}. The factor corrects the average of the frequencies of the mass multiplet less than \SI{4}{\%}. For both complexes the frequencies of the mass multiplets from experiment and DFT agree very well within \SI{1}{cm^{-1}}. This excellent agreement strongly supports the hypothesis that the measured vibrational modes are generated by the here presented \cnn and \ccn complexes. 

\begin{table}
\caption{\label{tab:phonon_cn2} Comparison of the asymmetric stretching mode frequency of neutral \cnn in different isotopic configurations between experiment (c.f. FIG. \ref{fig:isotope_shift}) and theory. The scaling factor is 0.9679. Numbers marked with $^*$ are from samples implanted with $^{15}$N in presence of $^{12}$C measured at 77 K (Ref. \onlinecite{Alt2015}). Numbers in brackets are the prediction for a possible experiment implanting $^{15}$N together with $^{13}$C.}
\begin{ruledtabular}
\begin{tabular}{ccc|ccc}
C & N$_1$ & N$_2$ & $\omega_{\text{calc}}$  &$\omega_{\text{scaled}}$ & $\omega_{\text{exp}}$  \\
  &   &   &   (cm$^{-1}$)  &  (cm$^{-1}$) &  (cm$^{-1}$)  \\
\hline
12 & 14 & 14 & 2128.3 & 2060.0 & 2059.6  \\
12 & 14 & 15 & 2118.0 & 2050.0 & 2050$^*$ \\
12 & 15 & 15 & 2107.4 & 2039.8 & 2040$^*$ \\
13 & 14 & 14 & 2069.7 & 2003.3 & 2003.7  \\
13 & 14 & 15 & 2059.1 & 1993.0 & (1993)  \\
13 & 15 & 15 & 2048.1 & 1982.4 & (1982)  \\
\end{tabular}
\end{ruledtabular}
\end{table}

%In addition, TAB. \ref{tab:phonon_cn2} contains a prediction for absorption band frequencies for a possible spectroscopic experiment enriching the only \SI{0.4}{\%} naturally abundant $^{15}$N, which are invisible in the experimental results from FIG. \ref{fig:isotope_shift}.

\begin{table}
\caption{\label{tab:phonon_c2n}  Comparison of the asymmetric stretching mode frequency of neutral \ccn in different isotopic configurations between experiment (see FIG. \ref{fig:isotope_shift}) and theory. The scaling factor is 0.9629}
\begin{ruledtabular}
\begin{tabular}{ccc|ccc}
C$_1$ & C$_2$ & N & $\omega_{\text{calc}}$  &$\omega_{\text{scaled}}$ & $\omega_{\text{exp}}$  \\
  &   &   &   (cm$^{-1}$)  &  (cm$^{-1}$) &  (cm$^{-1}$)  \\
\hline
12 & 12 & 14 & 2049.2 & 1973.1 & 1972.7  \\
12 & 13 & 14 & 2034.9 & 1959.3 & 1958.4  \\
13 & 12 & 14 & 1994.5 & 1920.4 & 1921.4 \\
13 & 13 & 14 & 1979.7 & 1906.2 & 1906.5  \\
\end{tabular}
\end{ruledtabular}
\end{table}

\subsection{Kohn-Sham molecular-like orbitals and bonding to the lattice}
\label{subsec:theo_molecular}

The calculated weak dependence of the asymmetric stretching mode on the supercell charge is comprehensible when the added or subtracted electrons do not occupy energy levels relevant for the bonds in the carbon-nitrogen complex. On the other hand, missing this dependence, the charge of the complexes cannot be deduced by simple comparison between measured and calculated vibrational frequencies.

To confirm this observation and to identify the charge of the complexes we further investigated the electronic structure. Since LDA is known to produce an inadequate picture of Kohn-Sham (KS) eigenvalues and especially of band gaps, we used the HSE06 hybrid functional to find the minimum of energy with atomic positions fixed at the final LDA positions. It turned out that a few of the KS orbitals are spatially localized around the complexes and resemble molecular-like orbitals of a free N-C-N and a C-C-N molecule. KS orbitals of the free molecules can be found in the Supplementary Material in FIG. S1 and S2.

For the neutral N-C-N free molecule with 14 valence electrons molecular-orbital theory predicts, analogous to the case of the CO$_2$, the orbitals $(1\sigma_g)^2$, $(1\sigma_u)^2$, $(2\sigma_g)^2$, $(1\pi_u)^4$, $(2\sigma_u)^2$, and $(1\pi_g)^2$ (partially occupied) as bonding and non-bonding, followed by further unoccupied antibonding orbitals. On the other hand, neutral C-C-N free molecule with 13 valence electrons gives rise to similar orbitals labeled with $\sigma$ and $\pi$. Since the C-C-N molecule has a linear, internuclear axis and the inversion symmetry is missing, no parity assignment are necessary. To compare localized KS molecular-like orbitals of \cnn with \ccn and with KS molecular orbitals from the free N-C-N and C-C-N molecules, we rename the above orbitals to $1\sigma$, $2\sigma$, $3\sigma$, $1\pi$, $4\sigma$, and $2\pi$. The integer labels the energy in ascending order. The electron occupation in the crystal is not partial as in the free molecule but turns out to be maximal, adopting the electrons from the surrounding Ga. Not considered are the six 1s core electrons heavily bound in three $\sigma$ orbitals.

The orbitals strongly localized around the complexes were identified by selecting the KS orbitals with sufficiently large localization on a C or N atom measured by means of the Mulliken charges ($q > 0.1$). Furthermore, the KS orbitals of the free complexes are taken into account for the identification. The $|\psi|= 0.05$ and \SI{0.1}{a_0^{-3/2}} isosurfaces for \cnn and \ccn are shown in FIG. \ref{fig:molecularorbitals} with the associated energies and charges of the participating atoms in TAB. \ref{tab:orbital}. The KS orbitals of the free N-C-N and C-C-N molecules are shown in the Supplementary Material in FIG. S1 and S2.

The localized orbitals of the complexes in GaAs are very similar to the orbitals of the molecule but extent for some eigenstates over the neighbouring Ga atoms, indicating the order of magnitude of bonding to the surrounding GaAs crystal. A Mulliken charge analysis helps to identify the bond structure inside the complexes and is depicted in TAB. \ref{tab:orbital}. Note that the charges of the lower orbitals sum up to nearly one and the charges of the higher orbitals to less than one, indicating an increasing delocalization of the KS orbitals and vice versa. The Mulliken charges for \ccn show a significant asymmetry resulting from the stronger electronegativity of N compared to C. Among these orbitals, the largest charge at the neighbouring Ga atoms is for the $1\pi$ orbital. It is suspected that this orbital gives the largest contribution to the bond between the N-C-N and C-C-N complex and the crystal.

\begin{table}
\caption{\label{tab:orbital} Localized molecular-like orbitals of \cnn and \ccn, their energy relative to the VBM and the assignment of charges to the two gallium atoms $q_\text{2Ga}$, the nitrogen atom $q_\text{N}$ and the carbon atom $q_\text{C}$ according to the Mulliken analysis. Note, the splitting of the degenerated $\pi$ orbitals for \ccn case.}
\begin{ruledtabular}
\begin{tabular}{cc|ccccc}
%\cnn &&&&&\\
\cnn &  $E$ (eV) & $q_\text{2Ga}$ & $q_\text{N}$ & $q_\text{C}$ & $q_\text{N}$ & $q_\text{2Ga}$ \\
\hline
$2\pi^4$     &  -5.8 & 0.01 & 0.19 & 0.09 & 0.01 & 0.16\\
$4\sigma^2$  &  -8.4 & 0.07 & 0.26 & 0.11 & 0.26 & 0.07\\
$1\pi^4$     &  -8.9 & 0.14 & 0.34 & 0.23 & 0.09 & 0.00\\
$3\sigma^2$  &  -9.3 & 0.06 & 0.30 & 0.10 & 0.30 & 0.06\\
$2\sigma^2$  & -20.5 &  0.07 & 0.37 & 0.10 & 0.37 & 0.07\\
$1\sigma^2$  & -22.6 &  0.01 & 0.29 & 0.41 & 0.29 & 0.01\\
\hline
\hline
%\ccn &&&&&\\
\ccn &  $E$ (eV) & $q_\text{2Ga}$ & $q_\text{N}$ & $q_\text{C}$ & $q_\text{C}$ & $q_\text{2Ga}$ \\
\hline
$2\pi^2$     &  -5.2 & 0.16 & 0.05 & 0.10 & 0.05 & 0.05 \\
$2\pi^2$     &  -5.5 & 0.02 & 0.26 & 0.13 & 0.00 & 0.13 \\
$4\sigma^2$  &  -7.5 & 0.01 & 0.01 & 0.02 & 0.06  & 0.08 \\
$1\pi^2$     &  -7.9 & 0.00 & 0.07 & 0.02 & 0.18 & 0.22 \\
$1\pi^2$     &  -8.6 & 0.21 & 0.35 & 0.14 & 0.02 & 0.00 \\
$3\sigma^2$  &  -8.6 & 0.14 & 0.49 & 0.10 & 0.05 & 0.03 \\
$2\sigma^2$  &  -15.5 & 0.03 & 0.06 & 0.25 & 0.40 & 0.22 \\
$1\sigma^2$  &  -21.2 & 0.10 & 0.62 & 0.29 & -0.01 & -0.01 \\
\end{tabular}
\end{ruledtabular}
\end{table}

\begin{figure}
\includegraphics[width=0.45\textwidth]{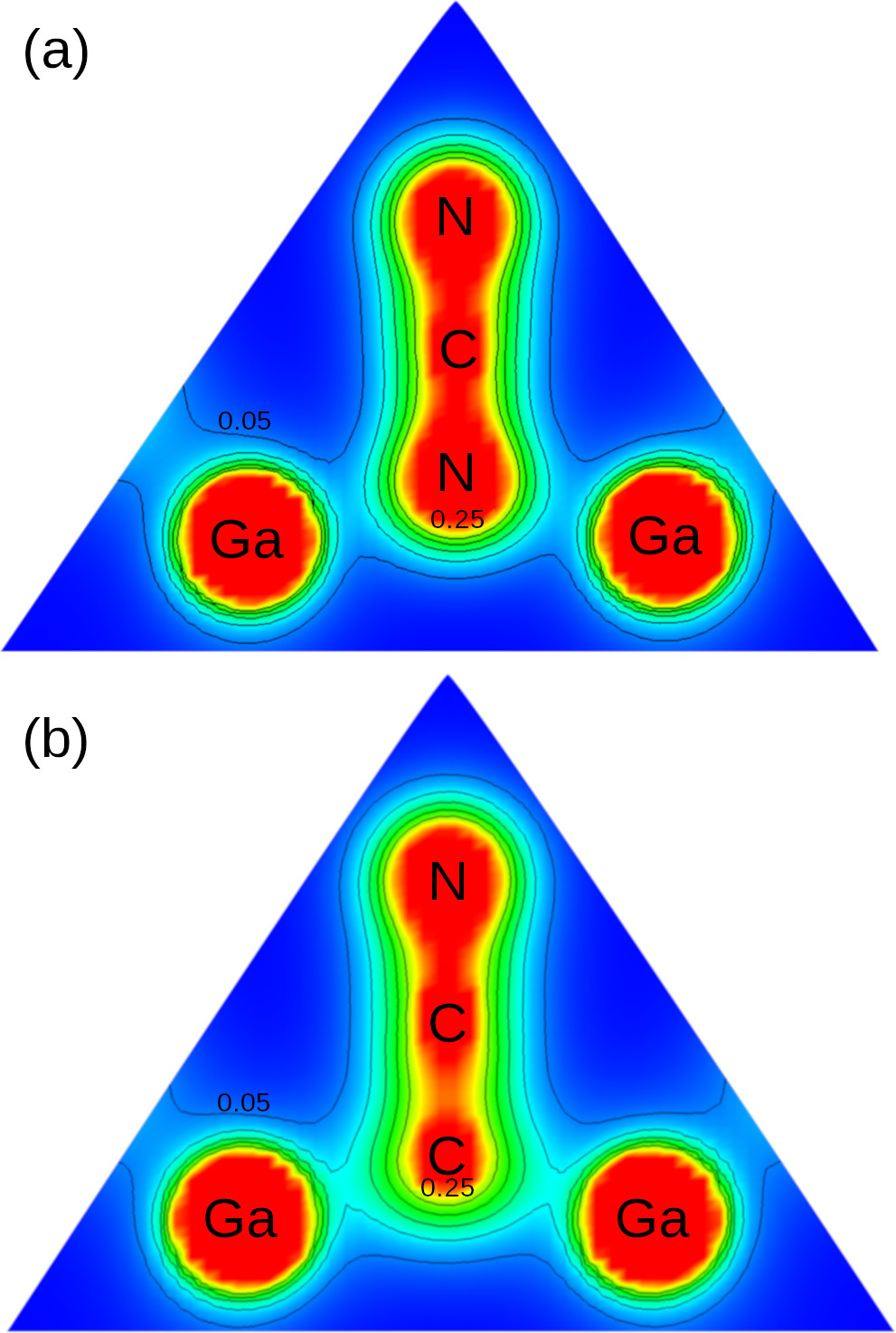}
\caption{\label{fig:tota_density} Full electron density of \cnn (a) and \ccn (b) in units of \si{\elementarycharge/a_0^3}. The values of the contour lines are equidistant beginning with \SI{0.05}{\elementarycharge/a_0^{3}} and ending with \SI{0.25}{\elementarycharge/a_0^{3}}.}
\end{figure}

From the Mulliken charge distribution, the strength of a bond cannot be unambiguously concluded. To support the statement about the $1\pi$ orbital as the major bond from the carbon-nitrogen complex to the surrounding GaAs crystal, a 2D section of the full electron density around the complex is shown in FIG. \ref{fig:tota_density}. The 2D section of (a) is aligned to the Ga-N-Ga plane and of (b) to the Ga-C-Ga plane. In comparison to the the single orbitals, the full electron density includes all orbitals and thus is more suitable for bond strength analysis. Since the electron density between C-N and C-C is much higher than the electron density to the surrounding Ga atoms in FIGs. \ref{fig:tota_density} (a) and (b), it can be concluded that the carbon-nitrogen complex can vibrate largely independently from the lattice, leading to the characteristics of a molecular-like LVM.

Furthermore, a closer inspection of the electron density in FIG. \ref{fig:tota_density} reveals the Ga-C bond to be stronger than the Ga-N bond. A chemical explanation may be, the smaller electronegativity of the C compared to the N which allows the Ga to collect more electron density for the Ga-C bond than for the Ga-N bond. Consequently, the C-C-N entity is overall stronger bonded to the crystal than the N-C-N entity. 

The above analyses of the molecular-like orbitals and the resulting bonding situation in the GaAs host lattice now allow a qualitative interpretation of the piezospectroscopic results. The linear N-C-N and C-C-N entities form valence bonds with the lattice only between the end atoms (N or C) and the (two) nearest-neighbour Ga atoms. The bond angles in the neutral charge state amount to \SI{147.5}{\degree} for \cnn (Ga-N-Ga) and \SI{144.9}{\degree} for \ccn (both Ga-C-Ga and Ga-N-Ga). In the sense of the discussion above and the charge-density result plotted in FIG. \ref{fig:tota_density}, we can treat the complexes approximately as a rigid unit. When applying stress in a \hkl<100> direction, the lattice is compressed in this direction and the bond angles become larger for those complexes oriented parallel to the stress direction. Therefore, the supporting action of the bonds decreases, inducing a slight downshift of the asymmetric stretching mode frequency. This is the reason for the negative sign of the piezospectroscopic parameter A$_1$ (see TAB. \ref{tab:uniaxial_stress}). On the other hand, for those complexes oriented perpendicular to the stress direction, bond angles will decrease and the frequency increase. Consequently, the parameter A$_2$ will be positive.

\subsection{Formation energies}
\label{subsec:theo_formation}

\begin{figure}
\includegraphics{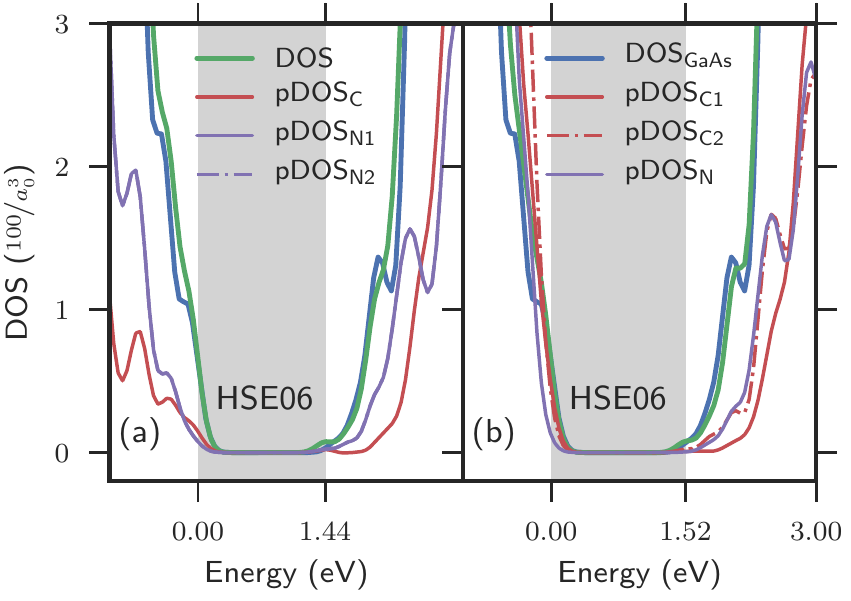}
\caption{\label{fig:densityofstates} Density of states (DOS) and the projected density of states (pDOS) of the C (pDOS$_\text{C}$) and N (pDOS$_\text{N}$) atom of the complex from HSE06 functional calculations are portrayed for the \cnn and \ccn complex in (a) and (b), respectively. DOS$_\text{GaAs}$ is the density of states for the pure GaAs host crystal. The DOS and DOS$_\text{GaAs}$ are scaled by $\nicefrac{1}{200}$}
\end{figure}

After clarifying the charge environment of the complexes, and the strength and positions of the bonds, the actual charge state of the complexes remains to be discussed. This property is determined from the energy levels inside the band gap, close to valence and conduction band edges, and their occupation (Fermi level). To answer this question the use of the HSE06 hybrid functional for the electronic structure is essential because of the smaller model uncertainty compared to the LDA functional. The band gaps from LDA (HSE06) DFT calculations were \SI{0.49}{eV} (\SI{1.57}{eV}), \SI{0.42}{eV} (\SI{1.44}{eV}) and \SI{0.49}{eV} (\SI{1.52}{eV}) for GaAs, and the centers \cnn and \ccn, respectively. The experimental band gap of GaAs is \SI{\approx 1.52}{eV}. The LDA functional calculations highly underestimate the experimental band gap and the HSE06 calculations provide a value close to the experimental gap.

To find potential thermodynamic charge transition levels established by the \cnn or \ccn complexes, we have calculated the density of states (DOS) and the projected density of states (pDOS) of the C (pDOS$_\text{C}$) and N (pDOS$_\text{N}$) atoms. The energies of the DOS and pDOS in FIG. \ref{fig:densityofstates} are aligned to its relative VBM and show no defect levels in the band gap for both carbon-nitrogen complexes. Although this observation gives a hint, regarding the fact that the hybrid functionals often give good results for the electronic structure, it is not compelling. 

\begin{figure}
\includegraphics[width=0.5\textwidth]{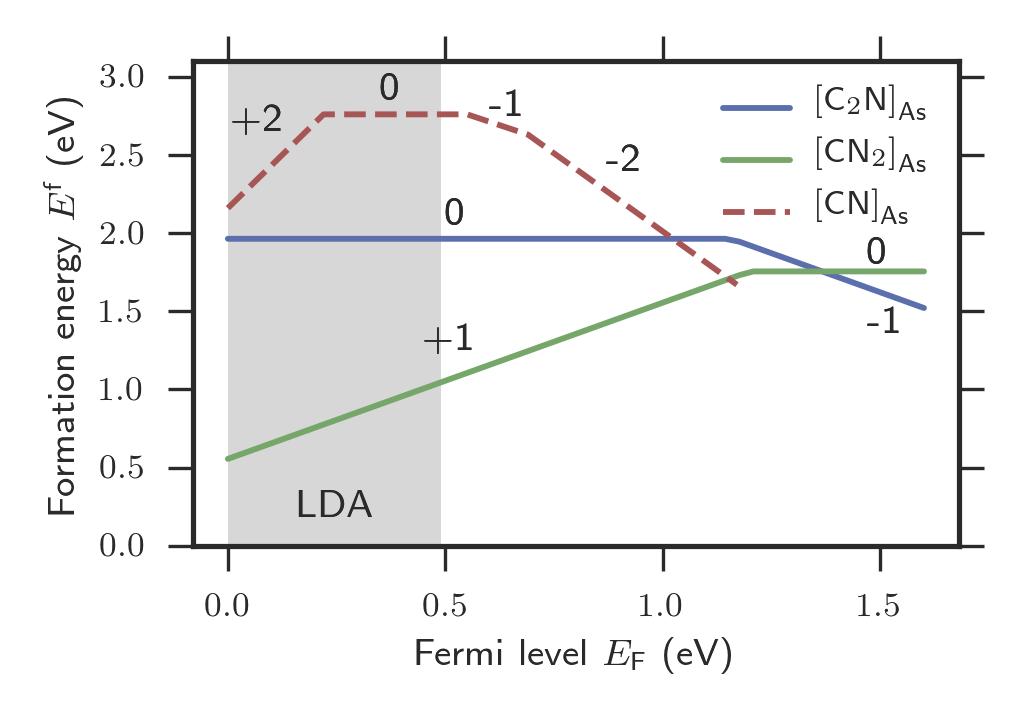}
\caption{\label{fig:formation_energies} Formation energies are calculated according to EQ. \ref{eq:formation_energy} except \cn which is from Ref. \onlinecite{Limpijumnong2008}. The gray shaded area indicates the band gap calculated with the LDA functional.}
\end{figure}

Therefore, we have calculated the formation energy as a function of the Fermi energy using the HSE06 hybrid functional according to EQ. \ref{eq:formation_energy}. Fig. \ref{fig:formation_energies} shows the Fermi energy dependent formation energies for both complexes in different charge states. Since only charge states with the lowest formation energy are drawn, the kinks indicate the thermodynamic charge transition levels. The charge transition levels were calculated using LDA structural energies and were aligned using the HSE06 VBM with respect to the average electrostatic potential according to Ref. \onlinecite{Ramprasad2012}. 

For comparison, we have inserted the formation energy of the proposed \cn complex as calculated by Limpijumnong et al. \cite{Limpijumnong2008}. In general, the formation energy of the \cnn and \ccn complex in GaAs are lower than the \cn complex, revealing them more likely. Furthermore, the formation energy of the \cnn complex is lower than of \ccn for almost the entire band gap. This coincides with the experimental observation that the vibrational frequency attributed for the \cnn center can be found in bulk GaAs whereas the signature of \ccn is only found in implanted samples. 

The charge transition levels $\epsilon\left(+1/0\right) = 1.18\,\text{eV}$ for \cnn and $\epsilon\left(0/-1 \right)= 1.15\,\text{eV}$ for \ccn are both located in a distance to the conduction band minimum (CBm) at \SI{1.52}{eV} and, therefore, could be interpreted as deep defects. On the other hand, for both the \cnn and \ccn complexes, no KS defect levels could be found inside the band gap, neither from LDA nor from HSE06 calculations, but only levels very close to the CBm (cf. FIG. \ref{fig:densityofstates}).

The results of both approaches to identify the thermodynamic charge transition levels and its associated charges are inconsistent on a certain level of accuracy. The DOS approach is limited because it considers KS orbitals, the formation energy approach is limited because it contains LDA energies and several correction terms. We conjecture that the charge transition level for both carbon-nitrogen complexes should actually move exactly to the CBm if calculated in a larger supercell and with even more accuracy provided by an still higher level of theory. When the charge transition levels are exactly at the CBm, the \ccn complex is neutral and the \cnn defect is positively charged for a regular Fermi energy.

\section{Conclusion}
The carbon-nitrogen complexes \cnn and \ccn form stable molecular-like complexes in GaAs. Piezospectroscopic FTIR investigations on the associated vibrational absorption bands at \SI{2060}{cm^-1} for \cnn and \SI{1973}{cm^-1} for \ccn prove that both complexes have tetragonal symmetry with the axis oriented parallel to the \hkl<100> direction.

The geometry prediction of the complexes from FTIR investigations was used to carry out DFT calculations. After geometry relaxation, the structure from DFT stayed in the predicted structure and, hence, substantiate the experimental investigation. Furthermore, vibrational frequencies of the highest and asymmetric stretching mode for both complexes are within a few wave numbers with the FTIR measurements including band shifts due to isotopes. 

The \cnn in GaAs complex is the structural analogue to the linear CO$_2$ molecule. In accordance with piezospectroscopic results, the substitutional position with the C atom on the anion site (As) and the N-C-N or C-C-N axis oriented along the \hkl<100> direction has lowest energy. The change of bond lengths and, consequently, the frequencies of the asymmetric stretching mode of both complexes with the charge state are relatively small.

The portrayed structural model for the complexes explains the negative piezospectroscopic parameter A$_1$ from the FTIR measurements. In addition, a detailed inspection of electronic charge density is consistent with the larger absolute value of A$_1$ of \ccn compared to \cnn.

Concluding the formation energies, the \cnn complex is more stable than the \ccn complex and a possible \cn complex is less likely. The most likely charge state assignment for \cnn is singly positive (1+) and for \ccn neutral (0). This results from an interpretation of the density of states and the formation energies.

Altogether, the consistency between experimental results and first-principles calculations in this study is excellent. 

%Conversely, regarding the quality which can be achieved in careful experiments and the accuracy which can be achieved with first principles calculations, an excellent agreement between both approaches should be required if a proposed defect structure is considered true.

\section*{Supplementary Material}
See supplementary material for the CN$_2$ and C$_2$N KS molecular-like orbitals from HSE06 DFT calculations.

\begin{acknowledgments}
The authors are indebted to W. Ulrici for helpful discussions and to U. Kretzer from Freiberger Compound Materials, Freiberg, Germany, for the supply of high-purity GaAs materials. The work was funded in part by the Deutsche Forschungsgemeinschaft (DFG). The authors gratefully acknowledge the Gauss Centre for Supercomputing e.V. (www.gauss-centre.eu) for funding this project by providing computing time on the GCS Supercomputer SuperMUC at Leibniz Supercomputing Center (LRZ, www.lrz.de).
\end{acknowledgments}

\bibliography{bib}% Produces the bibliography via BibTeX.

\end{document}